\documentclass[english,aps,twocolumn]{revtex4}
\usepackage{epsfig}
\usepackage{graphicx}
\usepackage{amsmath}

\makeatletter
\@ifundefined{textcolor}{}
{%
 \definecolor{BLACK}{gray}{0}
 \definecolor{WHITE}{gray}{1}
 \definecolor{RED}{rgb}{1,0,0}
 \definecolor{GREEN}{rgb}{0,1,0}
 \definecolor{BLUE}{rgb}{0,0,1}
 \definecolor{CYAN}{cmyk}{1,0,0,0}
 \definecolor{MAGENTA}{cmyk}{0,1,0,0}
 \definecolor{YELLOW}{cmyk}{0,0,1,0}
 }

\makeatother

\usepackage{babel}

\begin{document}

\title{Assessing the foundation of the Trojan Horse Method}

\author{C. A. Bertulani}

\affiliation{Department of Physics and Astronomy, Texas A\&M University-Commerce, Commerce, TX 75429-3011, USA}
\affiliation{Institut f\"{u}r Kernphysik, Technische Universit\"{a}t Darmstadt, Schlossgartenstra\ss{}e 9,  D-64289 Darmstadt, Germany}

\author{M. S. Hussein}

\affiliation{Instituto Tecnol\'{o}gico de Aeron\'{a}utica, DCTA, 12.228-900 S\~{a}o Jos\'{e} dos Campos, SP, Brazil}
\affiliation{Instituto de Estudos Avan\c{c}ados, Universidade de S\~{a}o Paulo C. P.
72012, 05508-970 S\~{a}o Paulo-SP, Brazil}
\affiliation{Instituto de F\'{\i}sica,
Universidade de S\~{a}o Paulo, C. P. 66318, 05314-970 S\~{a}o Paulo, SP, Brazil}

\author{S. Typel}

\affiliation{Institut f\"{u}r Kernphysik, Technische Universit\"{a}t Darmstadt, Schlossgartenstra\ss{}e 9,
  D-64289 Darmstadt, Germany}
\affiliation{GSI Helmholtzzentrum f\"{u}r Schwerionenforschung, Planckstra\ss{}e 1, D-64291 Darmstadt, Germany}

\keywords{Heavy ions, Breakup reactions, Trojan Horse Method}

\begin{abstract}
We discuss the foundation of the Trojan Horse Method (THM) within the Inclusive Non-Elastic  Breakup (INEB) theory.
We demonstrate that the direct part of the INEB cross section, which is of two-step character, becomes,
in the DWBA limit of the three-body theory with appropriate approximations and redefinitions, similar in structure
to the one-step THM cross section. We also discuss the connection of the THM to the Surrogate Method (SM), which
is a genuine two-step process.
\end{abstract}

\maketitle

\section{Introduction} 

The recent upsurge of interest in two-step nuclear reactions stems from two reasons: the application of indirect methods in reactor technology and nuclear astrophysics \cite{THM3}. The main aim is to extract cross sections of reactions of interest by studying more complex transfer reactions under favorable experimental conditions. In the first application, the surrogate method, $(d,p)$ reactions are employed to gain information on neutron induced compound reactions with $^{238}$U, $^{232}$Th and other nuclei in the actinide region, see, e.g.\ Refs. \cite{Burke:2006qu,Escher:2006xq,THM4,Escher:2012nda,Escher:2016nhz}.
The second application is in the field of nucleosynthesis of light and intermediate-mass nuclei during Big Bang and stellar evolution. Since the cross sections of these reactions at the astrophysical energies of interest, of a few keV's, are very small, one relies on the so-called Trojan Horse Method (THM) \cite{Bau86,Typel:2000,Spitaleri:2001vv,Baur2003,THM1,THM2,Piz15,Cherubini:2015sva,Spitaleri:2016hzy}. Within this method involving the reaction of a projectile, $a= b +x$, with a target, $A$, one is interested in the cross section $\sigma$ of, say, the direct rearrangement reaction
\begin{equation}
  x+ A \rightarrow y + B \: .
  \label{xA-yB}
\end{equation}
Then the THM cross section is written as
\begin{eqnarray}
  \lefteqn{\sigma_{\rm THM}(a+A \rightarrow b + y + B)}
  \nonumber \\
   & = & K_{\rm THM} \times |\phi(\textbf{k}_{b})|^{2}\times \sigma(x + A \rightarrow y +B) \: ,
\end{eqnarray}
where $K_{\rm THM}$ is a kinematic factor, $\phi$ is the momentum-space, internal wave function of the primary projectile, the Trojan horse $a$, and $\textbf{k}_b$ is the momentum of the spectator fragment, $b$. The merit of the THM resides in the premise that since $x$ is brought to the target position by the surrogate ion, $a$, most of the hindering effect of the Coulomb barrier is gone and the reaction (\ref{xA-yB}) proceeds more effectively above the Coulomb barrier. The other problem that complicates the measurement of the reaction (\ref{xA-yB}) at low energies for use in nuclear astrophysics is electron screening present if $x$ were a primary projectile \cite{BG10}. However the THM supplies a secondary $x$ projectile at above barrier energies, as explained above, and accordingly the electron screening problem is avoided. These conditions, no Coulomb barrier to surpass, and no electron screening,  allows the extraction of the desired cross section of reaction (\ref{xA-yB}) through the reaction
\begin{equation}
 a + A \rightarrow b + x + A \rightarrow b + y + B
\label{aA-byB}
\end{equation} 
with relative ease even at the extremely low energies required to simulate the conditions of the astrophysical environment. The THM has been very useful in supplying astrophysical S-factors of relevant reactions in different scenarios and energies at which the direct measurements are either not feasible or do not exist.

{\bf In this work we give the general structure of the inclusive non-elastic breakup cross section which is the basis of both the Trojan Horse Method and the Surrogate Method (SM) as argued in Ref.\ \cite{Hussein2017}. The motivation behind our use of the theory of inclusive breakup cross section as developed by \cite{IAV1985} and \cite{HM1985} is that it supplies a natural framework to investigate pieces of the cross section associated with particular processes. The THM aims at calculating a cross section associated with a direct process where the projectile bring in the desired fragment whose direct interaction with the target is sought for. The SM aims at calculating the cross section for a process where the projectile brings in a neutron or another fragment which subsequently forms a compound nucleus as it interacts with the target. All these processes are contained in the Inclusive Non-Elastic Breakup (INBU) cross section of the IAV theory \cite{IAV1985, Austern1987}, which we describe next.}

\section{Inclusive non-elastic breakup}

The THM deals with a cross section which is a part of the Inclusive Non-Elastic Breakup Cross Section (INEB) \cite{Hussein2017}. To exhibit this we recall the INEB cross section, \cite{CFH2017, HusseinFus2017},
\begin{equation}
\frac{d^{2}\sigma^{\rm INEB}_b}{dE_{b}d\Omega_{b}} = \hat{\sigma}_{R}^{x}  \ \rho_{b}(E_b), \label{inebI}
\end{equation}
where $\hat{\sigma}_{R}^{x}$ is the total reaction cross section of the interacting fragment, $x$, and 
\begin{equation}
\rho_{b}(E_b)  \equiv [d\textbf{k}_{b}/(2\pi)^3]/[dE_{b}d\Omega_{b}]  = \mu_{b}k_{b}/[(2\pi)^{3}\hbar^3]
\end{equation} 
is the density of state of the observed, spectator fragment, $b$. The reaction cross section $\hat{\sigma}_{R}^{x}$ is given by
\begin{equation}
\hat{\sigma}_{R}^{x} = - \frac{k_x}{E_x}\langle\hat{\rho}_{x}(\textbf{r}_x)\left|W_{x}(\textbf{r}_x)\right|\hat{\rho}_{x}(\textbf{r}_x)\rangle,
\label{inebII}
\end{equation}
where $W_{x}$ is the imaginary part of the complex optical potential, $U_{x}$, 
of the interacting fragment, $x$, in the field of the target, $A$.
The source function, $\hat{\rho}_{x}(\textbf{r}_x)$ is given by,
\begin{equation}
\hat{\rho}_{x}(\textbf{r}_x) = (\chi_{b}^{(-)}|\Psi^{(+)}_{3B}\rangle \: .
\end{equation}
The wave function, $|\Psi^{(+)}_{3B}\rangle$ is the exact three-body ($x + b + A$) wave function within the spectator model. Within the DWBA, this wave function acquires, in the post representation of Ichimura-Austern-Vincent (IAV) \cite{IAV1985},  the form 
\begin{equation}
 |\Psi^{(+)}_{3B}\rangle = (E - K_b - U_b - K_x - U_x + i\varepsilon)^{-1}
 V_{xb}|\phi_{a}\chi^{(+)}\rangle \: .
\end{equation}
In the prior form of the Udagawa-Tamura (UT) approach \cite{UT1981}, $V_{xb}$ is replaced by $(U_x + U_b - U_a)$. Accordingly, the source function $\hat{\rho}_{x}(\textbf{r}_x)$ becomes, in the post form,
\begin{equation}
\hat{\rho}^{\rm IAV}_{x}(\textbf{r}_x) 
= G_{x}^{(+)}(E_x)(\chi_{b}^{(-)}|V_{xb}|\phi_{a}\chi_{a}^{(+)}\rangle ,
\label{rhoIAV}
\end{equation}
and, in the prior form,
\begin{equation}
\hat{\rho}^{\rm UT}_{x}(\textbf{r}_x) 
= G_{x}^{(+)}(E_x)(\chi_{b}^{(-)}|(U_b + U_x - U_a)|\phi_{a}\chi_{a}^{(+)}\rangle,
\label{rhoUT}
\end{equation}
where $E_x = E - E_b$ and
\begin{equation}
 G_{x}^{(+)}(E_{x}) =  \left(E_{x} - K_x - U_x + i\varepsilon\right)^{-1}
\end{equation}
is the Green's function of particle $x$.
The connection between the two forms 
\begin{equation}
\hat{\rho}^{\rm IAV}_{x}(\textbf{r}_x) = \hat{\rho}^{\rm UT}_{x}(\textbf{r}_x) + \hat{\rho}^{\rm HM}_{x}(\textbf{r}_x),
\label{rhoIAVsumrule}
\end{equation}
is the non-orthogonality condition, or Hussein-McVoy (HM) source function \cite{Ichimura1990, HFM1990},
\begin{equation}
\hat{\rho}^{\rm HM}_{x}(\textbf{r}_x) = (\chi_b^{(-)}|\phi_{a}\chi_{a}^{(+)}\rangle \: .
\label{rhoHM}
\end{equation}

It has been verified that the INEB cross section (\ref{inebI}) calculated with the UT source function (\ref{rhoUT}) corresponds to the physical process of elastic breakup followed by fusion (capture) of $x$ with the target always remaining in the ground state. The full cross section calculated with the IAV source function (\ref{rhoIAV}) contains the UT term plus all other processes where the target is excited or other channels in the $x + A$ system are reached, accounted for by the HM contribution. Accordingly, we shall use the IAV description to discuss the nature of the THM. For this purpose we write for the imaginary part of the optical potential of the interacting fragment $x$, 
\begin{equation}
 \mbox{Im}U_x \equiv W_x = W^{D}_x + W^{CN}_x \: ,
\end{equation} 
where we have designated the direct processes in the $x + A$ system by $D$, and the compound nucleus processes by $CN$. Considering only the direct processes, we have, within the post form IAV theory, using the corresponding source function Eq. (\ref{rhoIAV}),
\begin{equation}
\frac{d^{2}\sigma^{\rm INEB, (D)}_b}{dE_{b}d\Omega_{b}} = - \rho_{b}(E_b) \frac{k_x}{E_x}\langle\hat{\rho}_{x}^{(+){\rm IAV}}|W^{D}_x|\hat{\rho}_{x}^{(+){\rm IAV}}\rangle.
\end{equation}
At this point we recall the general structure of $W_{x}^{D}$. If we call the projector of the direct non-elastic $x + A$ channels, $P^{(D)}_x$, and the elastic one by $P^{(0)}_x$, then (see Eqs. (\ref{A11}, {A18}) in the appendix \ref{App})
\begin{eqnarray}
- W_{x}^{D} &=& \pi P^{(0)}_{x}V P^{(D)}_{x}\delta(E_x - P^{(D)}_{x}H_{x}P^{(D)}_{x})P^{(D)}_{x}V P^{(0)}_{x} \nonumber \\
&=& \pi \sum_{f}\int \frac{d \textbf{k}_{f}}{(2\pi)^{3}} V_{(0,f)} |\chi_{f}^{(-)}(\textbf{k}_{f})\rangle\langle\chi_{f}^{(-)}(\textbf{k}_{f})| V_{(f,0)}\nonumber \\
& & \times\delta(E_x - E_f)  \label{w1}
\end{eqnarray}
with a sum over intermediate channels $f$ and integration over the corresponding momenta. We have simplified the notation through the introduction of 
$V_{0,f} \equiv P^{(0)}_{x}V P^{(D)}_{x}\equiv P^{(0)}_{x} H_{x}^{\rm (eff)}P^{(D)}_{x}$ upon the use of complete set of intermediate channels spanned by $P^{(D)}_{x}$, and used the spectral representation of the delta function as shown in the appendix \ref{App}.
Thus the structure of the direct part of INEB cross section becomes,
\begin{eqnarray}
\lefteqn{\frac{d^{2}\sigma^{\rm INEB, (D)}_b}{dE_{b}d\Omega_{b}} =\pi \rho_{b}(E_b) \frac{k_x}{E_x}} \nonumber \\
& & \times \sum_{f}\int \frac{d \textbf{k}_{f}}{(2\pi)^{3}}  \delta(E_x - E_f)
 |\langle\chi_{f}^{(-)}(\textbf{k}_{f})|V_{(0,f)}|\hat{\rho}_{x}^{(+){\rm  IAV}}\rangle|^2 .\nonumber \\
\label{sigmaIAV}
\end{eqnarray}
Consider one particular final $x + A$ channel, say $y + B$, then its contribution to the above cross section is
\begin{eqnarray}
\lefteqn{\frac{d^{4}\sigma^{\rm INEB, (D)}_{(b, y)}}{dE_{b}d\Omega_{b}dE_{y}d\Omega_{y}}} \nonumber \\
 & = &\pi\rho_{b}(E_b)\rho_{y}(E_y)\frac{k_x}{E_x}
|\langle\chi_{y}^{(-)}(\textbf{k}_{y})|V_{(x,y)}|\hat{\rho}_{x}^{(+){\rm IAV}}\rangle|^2 \nonumber \\
& = & \pi\rho_{b}(E_b)\rho_{y}(E_y)\frac{k_x}{E_x} \nonumber \\ & &
\times |\langle\chi_{y}^{(-)}(\textbf{k}_{y})|V_{(x,y)}G^{(+)}_{x}(\chi_{b}^{(-)}|V_{xb}|\phi_{a}\chi_{a}^{(+)}\rangle|^2 ,
\end{eqnarray}
using Eq.~(\ref{rhoIAV}).
Thus, within the spectator model, the cross section for the process
(\ref{aA-byB})
is described by an amplitude which is the product of the effective elastic breakup interaction, $(\chi_{b}^{(-)}|V_{xb}|\phi_{a}\chi_{a}^{(+)}\rangle$ (in the post representation) times the Green's function of the interacting fragment, $G^{(+)}_{x}(E_x)$, times the interaction $V_(x, y)$ for the transition (\ref{xA-yB}).
It is instructive to use an eikonal/Glauber-type approximation for the distorted wave of the projectile, 
\begin{equation}
 \chi^{(+)}_{a}(\textbf{r}_b, \textbf{r}_x) = \chi^{(+)}_{b}(\textbf{r}_b)\chi^{(+)}_{x}(\textbf{r}_x) \: .
\end{equation} 
Then defining the elastic breakup potential 
\begin{equation}
 V_{\rm (ebu)} \equiv \langle\chi^{(-)}_{b}|V_{xb}|\chi^{(+)}_{b}\phi_{a}\rangle \: ,
\end{equation}
the amplitude of the process involved in the THM has the structure
\begin{eqnarray}
\lefteqn{\langle\textbf{k}_{y}|T_{(a + A \rightarrow b + x + A \rightarrow b + y + B)}|\textbf{k}_{x}\rangle} \nonumber  \\
& = & \langle\chi^{(-)}_{y}|V_{(x, y)} G^{(+)}_{x}(E_x)V_{\rm (ebu)}|\chi^{(+)}_{x}\rangle,
\label{ampl-IAV}
\end{eqnarray}
clearly showing that the process, as described by the correct post-form IAV theory, is a two-step process.
Had we used the HM source function, Eq.(\ref{rhoHM}), we would have obtained for the 
amplitude of process (\ref{aA-byB}),
\begin{eqnarray}
\lefteqn{\langle\textbf{k}_{y}|T_{(a + A \rightarrow b + x + A \rightarrow b + y + B)}|\textbf{k}_{x}\rangle} \nonumber  \\
& = & \langle\chi^{(-)}_{y}V_{(x,y)}|(\chi^{(-)}_{b}|\chi^{(+)}_{b}\phi_{a}\chi^{(+)}_{x}\rangle \nonumber \\
& = & \langle\chi^{(-)}_{y}|V_{(x,y)}\hat{S}_{b}(\textbf{r}_x)|\chi_{x}^{(+)}\rangle,
\label{ampl-HM}
\end{eqnarray}
where the intrinsic  projectile wavefunction modified $b$-fragment elastic S-matrix element is given by 
\begin{eqnarray} 
 \hat{S}_{b}(\textbf{r}_x) & \equiv & 
\langle\chi^{(-)}_{(b,\textbf{k}^{\prime}_b)}(\textbf{r}_{b})|\phi_{a}(\textbf{r}_{b},
 \textbf{r}_{x})\chi^{(+)}_{(b,\textbf{k}_b)}(\textbf{r}_{b})\rangle \nonumber \\
 & = & \int d\textbf{r}_{b} \left[\chi_{(b, \textbf{k}^{\prime}_b)}^{(-)}(\textbf{r}_b)\right]^{\ast}\chi^{(+)}_{(b, \textbf{k}_{b})}(\textbf{r}_{b})\phi_{a}(\textbf{r}_b,\textbf{r}_x) \nonumber \\
& = & \int d\textbf{r}_{b} S_{(\textbf{k}^{\prime}_{b}, \textbf{k}_b)}(\textbf{r}_{b}) \phi_{a}(\textbf{r}_{b},\textbf{r}_x) \: .
\end{eqnarray}
The apparent one-step process of the HM version of the THM is quite clear!

Since we are considering the direct part of the inclusive non-elastic breakup, the UT process, which is a manifestly a compound nucleus (of the x + A system) process, does not contribute. Accordingly, the general IAV cross section which has the form, cf.\ Eq.~(\ref{rhoIAVsumrule}) and Eq.~(\ref{sigmaIAV}),
\begin{equation}
\sigma_{\rm (IAV)} = \sigma_{\rm (UT)} + \sigma_{\rm (HM)} + \sigma_{\rm (Interference)},
\end{equation}
where the interference term, which is a pseudo cross section as it can be negative is of the general structure $\sigma_{\rm (interference)} = 2\mbox{Re}[A^{\dagger} A]$, where 
$A$ is proportional to the ``expectation"  value $ \langle \hat{\rho}^{(+) {\rm UT}}|W_{x}|\hat{\rho}^{(+){\rm HM}}\rangle$.
The direct part of the cross section, however, is 
\begin{equation}
\sigma_{\rm (IAV, D)} = \sigma_{\rm (HM, D)}.
\end{equation}
Thus we reach the important result,
\begin{equation}
\sigma_{\rm (IAV, D, 2-step)} = \sigma_{\rm (HM, D, 1-step)}.
\label{major}
\end{equation}
Accordingly a two-step process ``collapses" into a one-step process!

In the HM one-step process the modified elastic S-matrix of $b$, $\hat{S}_{b}(\textbf{r}_x)$ appears in the amplitude Eq. (\ref{ampl-HM}) multiplying the interaction $V_{(x,y)}$ in the $x +A \rightarrow y + B$ amplitude. Therefore, the corresponding cross section will have the general form,
\begin{equation}
\frac{d^{2}\sigma_{\rm (HM)}}{dE_{b}d\Omega_{b}} = \rho_{b}(E_b) \frac{k_x}{E_x}|\langle\chi_{y}^{(-)}|V_{(x,y)}\hat{S}_{b}(\textbf{r}_x)|\chi_{x}^{(+)}\rangle|^2 .
\end{equation}
This is similar to the THM cross section if an effective $V^{\rm (eff)}_{(x, y)} = \hat{S}_{b}(\textbf{r}_x)V_{(x,y)}$ is introduced, and a further approximation is made with respect to the $b$-fragment modified elastic S-matrix $\hat{S}_{b}(\textbf{r}_x)\approx  \phi(\textbf{k}_b)$, implying a maximum survival probability of $b$. Then we recover the THM cross section 
\begin{eqnarray}
\frac{d^{2}\sigma_{\rm THM}}{dE_{b}d\Omega_{b}} & = & \rho_{b}(E_b) \frac{k_x}{E_x}|\langle\chi^{(-)}_{y}|V^{\rm (eff)}_ {(x, y)}|\chi^{(+)}_{x}\rangle|^2 \nonumber \\
& = & K_{\rm (THM)}|\phi(\textbf{k}_b)|^{2} \sigma_{(x +A \rightarrow y + B)}
\end{eqnarray}
as a product of a kinematic factor, the momentum distribution of the spectator, and a two-body reaction cross section.

\section{The THM as derived by Baur and collaborators}

In Ref.\ \cite{Bau86} Baur uses the post form of the DWBA to describe the desired cross section for the direct process at hand, (\ref{aA-byB}). 
\begin{equation}
T_{\rm (Baur)} = \langle\chi^{(-)}_{b,yB}\Psi^{(-)}_{yB}\phi_{b}|V_{bx}|\Psi^{(+)}_{aA}\phi_{a}\phi_{A}\rangle,
\end{equation}
where $\Psi^{(+)}_{aA}\equiv \Psi^{(+)}_{bxA}$ is the full three-body scattering wave function in the incident channel. To proceed further, in Ref.\ \cite{Baur2003}  the DWBA approximation was employed for this three-body wave function, $\Psi^{(+)}_{bxA}\approx \chi^{(+)}_{aA}$. With appropriate use of surface dominated $\Psi^{(-)}_{yB}$ which allows using its asymptotic form and set the interior of it to zero, one obtains Baur's one-step T-matrix. In a way this work which describes the THM as a one-step process, is similar to that of HM \cite{HM1985}. {\bf Instead of the direct use of the DWBA scattering wave function, if Ref.\ \cite{Baur2003} were to deal with the three-body scattering wave function in their treatment as Ref.\ \cite{Austern1987} has done, they would, in the DWBA-limit have obtained a two-step THM. Within the three-body approach of Ref.\ \cite{Austern1987}, the full three-body wave function of the initial channel is used, and its three Faddeev components are resolved, }
\begin{equation}
\Psi^{(+)}_{aA} = \Psi^{(+)}_{3B}\phi_{a}
\end{equation}
where,
\begin{equation}
\Psi^{(+)}_{3B} = \Psi^{(+)}_{xb} + \Psi^{(+)}_{xA} + \Psi^{(+)}_{bA} \: .
\end{equation} 
The dominant Faddeev component above has the general form
\begin{equation}
\Psi^{(+)}_{xb} = G_{0}^{(+)}V_{xb}\Psi^{(+)}_{3B} \: ,
\end{equation}
and thus the three-body wavefunction becomes,
\begin{equation}
\Psi^{(+)}_{3B} = G^{(+)}_{x,b}V_{xb}\Psi^{(+)}_{x,b} \: ,
\end{equation}
whose DWBA approximation is
\begin{equation}
\Psi^{(+)}_{3B} \approx G^{(+)}_{x,b}V_{xb}\chi^{(+)}_{a} \: .
\end{equation}
Thus the amplitude for the reaction (\ref{aA-byB}) is
\begin{equation}
T_{\rm (Baur)} = \langle\chi^{(-)}_{b,yB}\Psi^{(-)}_{yB}\phi_{b}|\left[V_{b, y}G^{(+)}_{x,b}V_{x,b}\right]|\chi^{(+)}_{a}\phi_{a}\rangle,
\end{equation}
with  
the optical, $x + b$ Green's function
\begin{equation}
  G^{(+)}_{x, b} = [E - K_b - K_x - U_x - U_b + i\varepsilon]^{-1} \: ,
\end{equation}
where the kinetic energy operators are denoted by $K$ and the optical potentials by $U$.
Luckily, as we have shown for the direct TH, $x + A \rightarrow y + B$, process, the direct DWBA two-step process is formally equal to the one-step process. Accordingly, we can use 
Eq. (\ref{major}) to write
\begin{equation}
\sigma_{\rm (IAV, D, 2-step)} = \sigma_{\rm (HM, D, 1-step)} \approx \sigma_{\rm (THM)}.
\end{equation}
Several approximations are required to show that $\sigma_{\rm (IAV, D, 2-step)} \rightarrow \sigma_{\rm THM}$,
and $\sigma_{\rm (HM, D, 1-step)} \rightarrow \sigma_{\rm THM}$. Clearly these approximations must be assessed in the future.

\section{Discussion, conclusions and outlook} 

In this Letter we have discussed the foundation of the Trojan Horse Method within the general framework of the Inclusive Non-Elastic Breakup reaction theory. We accomplished this by extracting the cross section for the exclusive process described by the THM.  This is made possible by inspecting the direct component of the imaginary part of the optical potential of the interacting fragment, $x$, in the inclusive reaction $a + A \rightarrow b + (x + A)$. The resulting cross section of the exclusive $a + A \rightarrow b + x + A \rightarrow b + y + B$ was found to be that of a one-step process. This is true as long as the THM is purely direct. By the same reasoning, in the Surrogate Method, if an exclusive reaction is considered proceeding through the compound nucleus, it is a genuine two-step process. It is important to check these findings by carefully considering the structure of these cross sections in the more general case of exotic nuclei. {\bf Our analysis of the THM allows for potential improvement of the method as one can trace the different steps used, within the INEB theory to reach the one-step THM. Knowing that the reaction purported to be described by the THM, $a + A \rightarrow b + x + A \rightarrow b + y + B$, (with the cross section of the subsystem $x + A \rightarrow y + B$ being the desired one), is a two-step process, one can improve the THM, without jeopardizing the numerical convenience of the one-step THM. Although not stated explicitly in our discussion above, the main difference between an amplitude describing a two-step direct process and that of a one-step process, is the presence of a propagator (Green function) in the former. In the context of the THM, this Green's function describes the intermediate propagation in elastic breakup channel, $b + x + A$. To reduce the Green's function into a factor, one uses the on-energy approximation and leave out the principal part. This automatically results in a one-step-like amplitude. Corrections  to the THM can then be made by an approximate treatment of this principal part.}

{\it Acknowledgements.} 
This work was partly supported by the US-NSF and by the Brazilian agencies, Funda\c c\~ao de Amparo \`a Pesquisa do Estado de
 S\~ao Paulo (FAPESP), the  Conselho Nacional de Desenvolvimento Cient\'ifico e Tecnol\'ogico  (CNPq). CAB also acknowledges a Visiting Professor support from FAPESP and MSH acknowledges a Senior Visiting Professorship granted by the Coordena\c c\~ao de Aperfei\c coamento de Pessoal de N\'ivel Superior (CAPES), through the CAPES/ITA-PVS program and support by the U.S. DOE grants DE-FG02-08ER41533 and the U.S. NSF Grant No. 1415656. ST was supported by the DFG through grant No.~SFB1245.

\appendix

\section{The reactive content of the imaginary part, $W_x$ of the $x + A$ optical potential}
\label{App}

To exhibit the detailed reactive content of $W_{x}$ we consider the $x + A$ scattering system. We introduce projection operators $P_x$ and $Q_x$, such that $P_x + Q_x = 1$, $P_{x}Q_{x} = Q_{x}P_{x} = 0$, and $P_{x}^2 = P_x$ and $Q_{x}^2 = Q_x$. $P_x$ is defined such that it projects out the open $x + A$ channels, while $Q_{x}$ projects out the closed, compound nucleus $x + A$ channels. Denoting the Hamiltonian that describes the dynamics of the $x + A$ subsystem by $H_x$, we have as usual, for the open channels (the direct reactions coupled equations)
\begin{equation}
(E_x - P_{x}H_{x}P_{x})P_{x}|\Psi^{(+)}\rangle = P_{x}H_{x}Q_{x}Q_{x}|\Psi_{xA}^{(+)}\rangle ,
\end{equation}
and for the closed, compound nucleus channels,
\begin{equation}
(E_x - Q_{x}H_{x}Q_{x})Q_{x}|\Psi_{xA}^{(+)}\rangle = Q_{x}H_{x}P_{x}|\Psi_{xA}^{(+)}\rangle .
\end{equation}
The closed channels equations above can be formally solved to give,
\begin{equation}
Q_{x}|\Psi_{xA}^{(+)} \rangle = \frac{1}{E_{x} - Q_{x}H_{x}Q_{x}}Q_{x}H_{x}P_{x}|\Psi_{xa}^{(+)}\rangle .
\end{equation}
Accordingly, we have for the open $x +A$ channels,
\begin{multline}
\Big(E_{x} - P_{x}H_{x}P_{x}  \\
- P_{x}H_{x}Q_{x}\frac{1}{E_{x}- Q_{x}H_{x}Q_{x}}Q_{x}H_{x}P_{x}\Big)P_{x}|\Psi_{xA}^{(+)} \rangle= 0.
\end{multline}
The above equation is exact and useless! The $Q_x$ propagator, 
$(E_{x} - Q_{x}H_{x}Q_{x})^{-1}$, has poles whenever a compound $x + A$ resonance is excited. This very strong energy dependence is removed by introducing the energy average $Q_{x}$ propagator, $(E_{x} - Q_{x}H_{x}Q_{x} + iI)^{-1}$, where $I$ is a large energy 
that encompasses many $x +A$ compound nucleus resonances. Thus calling the optical $x + A$ open channels wave function by $|\overline{\Psi}_{xA}^{(+)}\rangle$,  and the effective, complex, optical potential that accounts for the coupling to the closed $x + A$ compound nucleus channels, by $U_{x}^{(CN)}$, we have
\begin{eqnarray}
 \lefteqn{(E_{x} - P_{x}H_{x}P_{x} - U_{x}^{(CN)})|\overline{\Psi}_{xA}^{(+)}\rangle}
 \nonumber \\
 & \equiv & (E_{x} - P_{x}H_{x}^{\rm (eff)}P_{x})|\overline{\Psi}_{xA}^{(+)}\rangle ,
\end{eqnarray}
where we have introduced the effective $P_x$ projected Hamiltonian $H_{x}^{\rm (eff)} = H_{x} + U_{x}^{(CN)}$.
At this point we split the projection operator $P_x$ into the elastic $x + A$ channel projector (which corresponds to the elastic breakup channel in the full $b + x + A$ system), 
$P_{x}^{0}$, and the projector onto all the open non-elastic, direct channels, $P_{x}^{D}$. The resulting coupled channels equations are,
\begin{eqnarray}
 \lefteqn{(E_x - P_{x}^{(0)}H_{x}^{\rm (eff)}P_{x}^{(0)})P_{x}^{(0)}|\overline{\Psi}_{xA}\rangle} \nonumber \\
&  = & \left[P_{x}^{(0)}H_{x}^{\rm (eff)}P_{x}^{(D)}\right]P_{x}^{(D)}|\overline{\Psi}_{xA}\rangle
\label{direct}
\end{eqnarray}
and
\begin{eqnarray}
 \lefteqn{(E_x - P_{x}^{(D)}H_{x}^{\rm (eff)}P_{x}^{(D)})P_{x}^{(D)}|\overline{\Psi}_{xA}\rangle} \nonumber \\
& = & \left[P_{x}^{(D)}H_{x}^{\rm (eff)}P_{x}^{(0)}\right]P_{x}^{(0)}|\overline{\Psi}_{xA}\rangle \: .
\end{eqnarray}
Solving for the non-elastic direct channels, $P_{x}^{(D)}|\overline{\Psi}_{xA}\rangle$, we obtain for Eq.~(\ref{direct})
\begin{eqnarray}
 \lefteqn{\Big(E_x - P_{x}^{(0)}H_{x}^{\rm (eff)}P_{x}^{(0)}} \nonumber    \\ 
 & & - P_{x}^{(0)}H_{x}^{\rm (eff)}P_{x}^{(D)}G_{x}^{(+), D}P_{x}^{(D)}H_{x}^{\rm (eff)}P_{x}^{(0)}\Big)P_{x}^{(0)}|\overline{\Psi}_{xA}\rangle \nonumber \\
 & = & 0 
 \end{eqnarray}
after defining the Green's function
\begin{equation}
G_{x}^{(+), D}\equiv [E_{x} - P_{x}^{(D)}H_{x}^{\rm (eff)}P_{x}^{(D)} + i\varepsilon]^{-1} \: .
\end{equation}
The imaginary part of the average optical potential of the $x$ fragment is thus 
\begin{eqnarray}
-W_{x} & = & \mbox{Im}\left[H_{x}^{\rm (eff)} + H_{x}^{\rm (eff)}P_{x}^{(D)}G_{x}^{(+), D}P_{x}^{(D)}H_{x}^{\rm (eff)}\right] \: . \nonumber \\ & & 
\end{eqnarray}

We are now in a position to analyze the reactive content of $W_x$. Clearly, $\mbox{Im}[H_{x}^{\rm (eff)}]$ is the compound nucleus absorption contribution. We call this $W_{x}^{(CN)}$. The second contribution, $\mbox{Im}[H_{x}^{\rm (eff)}P_{x}^{(D)}G_{x}^{(+), D}P_{x}^{(D)}H_{x}^{\rm (eff)}]$, accounts for the direct non-elastic absorption, $W_{x}^{(D)}$. Note that the couplings 
$P_{x}^{(0)}H_{x}^{\rm (eff)}P_{x}^{(D)}$ are complex owing to the complexity of $H_{x}^{\rm (eff)}$. However, we make the assumption the compound nucleus absorption (fusion) has only diagonal couplings. Accordingly, 
\begin{equation}
-W_{x}^{(D)}  \approx  H_{x}^{\rm (eff)}P_{x}^{(D)}\mbox{Im}[G_{x}^{(+), D}]P_{x}^{(D)}H_{x}^{\rm (eff)} \: .  \label{A11} 
\end{equation}
The imaginary part of the Green's function $G_{x}^{(+), D}$ is
\begin{equation}
\mbox{Im}[G_{x}^{(+), D}] = \mbox{Im}\left[\frac{1}{E_x - P_{x}^{(D)}H_{x}^{\rm (eff)}P_{x}^{(D)} + i\varepsilon}\right]
\end{equation}
where the $P_{x}^{(D)}$ projected effective Hamiltonian is non-Hermitian. Thus the imaginary part of the Green's function must involve this feature. 
To simplify the notation we call the $P_{x}^{(D)}$ projected effective Hamiltonian, $K_0 + V$, where $K_0$ is the kinetic energy operator in the $P_{x}^{(D)}$ projected space. Then, following Ref.~\cite{CH2013},
\begin{eqnarray}
\mbox{Im} G(z) & = & \mbox{Im}\left[\frac{1}{z - K_0 - V + i\varepsilon}\right] \nonumber \\
& = & -\pi[1 + G^{(-)} V^{\dagger}]\delta (z - K_0)[1 + V(G^{(-)})^{\dagger}] 
 \nonumber \\ & & - (G^{(+)})^{\dagger} \mbox{Im} [V] G^{(+)},
\end{eqnarray}
where
\begin{equation}
G^{(-)} = \left[z - K_0 - V^{\dagger} - i\varepsilon\right]^{-1} \: ,
\end{equation}
and the M\"{o}ller operator $[1 + G^{(-)}V^{\dagger}]$ generates a distorted wave when operating on a plane wave,
or more generally a Coulomb distorted wave when $K_0$ is replaced by the Coulomb Hamiltonian $K_0 + V_{C}(r)$.
Using the spectral decomposition of the delta function, 
\begin{equation}
 \delta(z - K_0) = \int \frac{d\textbf{k}}{(2\pi)^{3}} \:
  |\textbf{k}\rangle \delta(z - E_k)\langle\textbf{k}| \: ,
\end{equation} 
then $\mbox{Im}G$ becomes
\begin{eqnarray}
\mbox{Im} G(z) & = & -\pi \int \frac{d\textbf{k}}{(2\pi)^3}|\chi_{\textbf{k}}^{(-)}\rangle \delta(z - E_k)\langle \chi_{\textbf{k}}^{(-)}| \nonumber \\
 & &  - (G^{(+)})^{\dagger}\mbox{Im} [V] G^{(+)} \: .
\end{eqnarray}
The above identity, when used in the context of $\mbox{Im}[G_{x}^{(+), D}]$, would give 
\begin{eqnarray}
 \mbox{Im}[G_{x}^{(+), D}] & = & 
-\pi \sum_{D}\int \frac{d \textbf{k}_{D}}{(2\pi)^{3}}  \: 
|\chi_{{\textbf{k}_D}}^{(-)}\rangle \delta(E_x - E_{{k_D}})\langle \chi_{{\textbf{k}_D}}^{(-)}| \nonumber \\
& & - [G_{x}^{(+), D}]^{\dagger}\mbox{Im}[P_{x}^{(D)}H_{x}^{\rm (eff)}P_{x}^{(D)}]G_{x}^{(+), D} \: .\nonumber \\
 \end{eqnarray} 
The last term in the above equation represents, when used in the cross section formula,  compound nucleus (fusion) coupling effects in the non-elastic direct channels projected out by $P_{x}^{(D)}$. These processes are of the type $x + A \rightarrow c + C \rightarrow CN$, and accordingly we ignore them. Thus the final form of $\mbox{Im}[G_{x}^{(+), D}]$, used in the derivation of Eq. (\ref{w1}), is
\begin{equation}
 \mbox{Im}[G_{x}^{(+), D}] = -\pi \sum_{D}\int \frac{d \textbf{k}_{D}}{(2\pi)^{3}}
|\chi_{{\textbf{k}_D}}^{(-)}\rangle \delta(E_x - E_{{k_D}})\langle \chi_{{\textbf{k}_D}}^{(-)}| \: . 
\label{A18}
\end{equation}

\end{document}